\documentclass[letterpaper,twocolumn,prb,amsmath,amssymb]{revtex4-1}

\usepackage{amsmath}
\usepackage{amssymb}
\usepackage{amsfonts}
\usepackage{graphicx}
\usepackage{dcolumn}
\usepackage{float}
\usepackage{bm}
\usepackage{mathrsfs}
\usepackage{txfonts}
\usepackage{tabularx}
\usepackage{bigstrut}
\usepackage[usenames,dvipsnames]{color}

\begin{document}

\title{\Large All-optical injection of charge, spin and valley currents in monolayer transition metal dichalcogenides }

\author{Rodrigo A. Muniz and J. E. Sipe }

\affiliation{Department of Physics and Institute for Optical Sciences, University of Toronto, Toronto ON, M5S 1A7, Canada }

\date{\today}

\begin{abstract}
Monolayer transition metal dichalcogenides have recently become a
playground for spin- and valleytronics research. Their low energy
spectrum can be described by Dirac cones on the corners of Brillouin
zone, but the physical properties are richer than those of graphene since the spin degeneracy is
lifted and the optical selection rules are valley dependent. This has been exploited for the optical injection of spin and valley polarized currents by the application of static electric fields.
In this paper we consider an all-optical method for the injection of charge, spin
and valley polarized currents. 
The presence of both a fundamental optical field and
its second harmonic can lead to the injection of currents due to
a nonlinear effect involving the quantum interference between one-
and two-photon absorption processes. We analyze how the injected quantities
can be controlled through the parameters of the incident light fields, allowing capabilities of control beyond those achieved with static fields, and discuss the conditions for experimental verification of our
results. 
\end{abstract}

\maketitle

\section{Introduction }

The optoelectronic properties of two dimensional materials are often qualitatively different from those of their three dimensional counterparts.
Novel fields of research that arise from these differences, such
as valleytronics \cite{butler13}, hold promise for the development of new technologies.
In addition, these materials can be conspicuously integrated into heterostructures, as coatings for example, paving the way for their application in integrated optical devices. 
An outstanding class of two dimensional materials are the monolayer transition metal dichalcogenides (TMD), which have recently been shown to
display several interesting electronic and optical properties \cite{wang12, xu14}.
Their atomic structure consists of an hexagonal lattice, and the low
energy spectrum is described by gapped Dirac cones on the corners
of the Brillouin zone. Due to strong spin-orbit coupling and broken
inversion symmetry, the spin degeneracy is lifted in opposite ways
in the two valleys, and the optical selection rules are valley dependent
\cite{xiao12,rostami13}. Therefore, optically excited carriers are
valley polarized and, for low enough photon energies, also spin polarized both according to the helicity of incident light
\cite{carbotte12,rose13,niu13}. The injected carriers can be then driven by an electric field, providing a valley and spin polarized current \cite{mak12,zeng12,jones13}.
Such currents have been the subject of intensive research, with respect to both fundamental questions and technological applications \cite{wang12}. 

Even though some nonlinear optical properties of monolayer TMDs have been studied \cite{kumar13,malard13}, the proposals for current injection have so far focused on the application of static fields. 
However, the need of a static applied electric field does not allow for fast switching, and offers only limited control of the currents.
It would be desirable to have an all-optical method for the injection of currents, since
it would allow for faster switching and more refined control over
the quantities of interest by using, for example, the polarizations and phase parameters of the incident fields.

Effective all-optical injection of currents can be achieved by coherent control. It makes use of both a fundamental optical field and its second harmonic, which allows for optical injection of currents by a nonlinear process involving quantum interference between one- and two-photon absorption \cite{rioux12}. 
It has been applied in several
experimental scenarios involving bulk and nanostructure semiconductors \cite{rioux12, bhat05, rumyantsev06, reiter09, rioux10, kiran11, virk11}, it has been predicted and seen in graphene \cite{norris10,rioux11,kiran12}, 
and experiments to lead to its observation in topological insulators have recently been proposed \cite{muniz14}. 
Here we study how it can be used for the injection
of polarized currents in monolayer TMDs. We compute the optical injection
rates of several quantities for monolayer films of TMDs. The quantities
considered are carrier, spin and valley polarization densities, as
well as charge, spin and valley currents. We show how the polarization
and a relative phase parameter of the incident fields can be used to control the
optical generation of quasiparticles and their currents on time scales set by the duration of laser pulses.

This article is organized in the following form: in Sec. \ref{sec:mos2}
we present the model for monolayer TMDs used for our calculations.
Sec. \ref{sec:inject} contains an outline of the computation for
the optical injection rate coefficients corresponding to carrier,
spin and valley densities, as well as charge, spin and valley currents.
In Sec. \ref{sec:results} we show the results obtained for different
polarizations and relative orientations of the incident fields. The
explicit expression for the injection rate coefficients are shown
in the Appendix \ref{app:coeffs}. We conclude with a discussion about
the experimental verification of our results in Sec. \ref{sec:discussion}.
Since the experimental techniques required to confirm our results
are well established, we expect that such experiments will help advance
the understanding and applications of optically injected currents
in monolayer TMDs.

\section{Model for monolayers of transition metal dichalcogenides }

\label{sec:mos2}

The computation of injection rates is performed using Fermi's Golden Rule in a method described earlier \cite{muniz14}, where general
expressions were provided for a two-band Hamiltonian.

The simplest model for TMDs has a 4-band Hamiltonian with 2-valleys
that for each lattice momentum $\boldsymbol{k}$ can be represented
by the matrix \cite{xiao12,rostami13} 
\begin{equation}
\hbar^{-1}H_{\tau s,\boldsymbol{k}}=t\left(\tau k_{x}\sigma_{x}+k_{y}\sigma_{y}\right)+\frac{\Delta}{2}\sigma_{z}+\frac{\lambda\tau s}{2}\left(\sigma_{0}-\sigma_{z}\right),
\label{eq:Hamilt}
\end{equation}
where $\Delta$ and $\lambda$ are parameters with dimensions
of frequency, $t$ is another parameter with dimension of velocity, $\tau=\pm1$ is the valley index, indicating either the $K$
($+1$) or $K^{\prime}$ ($-1$) point; and $s=\pm1$ is the spin-$\hat{z}$
index, indicating either $\left(\uparrow\right)$ or $\left(\downarrow\right)$
spin. The Hamiltonian \eqref{eq:Hamilt} splits into four $2\times2$ orthogonal sectors and can be
described in the generic form of a two-band system Hamiltonian 
\begin{equation}
H_{\boldsymbol{k}}=\hbar\varpi_{\boldsymbol{k}}\sigma_{0}+\hbar\boldsymbol{d}_{\boldsymbol{k}}\cdot\boldsymbol{\sigma}
\end{equation}
with 
\begin{equation}
\begin{array}{rl}
\varpi_{\boldsymbol{k}}= & \frac{\lambda\tau s}{2},\\
\boldsymbol{d}_{\boldsymbol{k}}= & t\tau k_{x}\hat{\boldsymbol{x}}+tk_{y}\hat{\boldsymbol{y}}+\Delta_{\tau s}\hat{\boldsymbol{z}},
\end{array}
\end{equation}
where $\Delta_{\tau s}=\left(\Delta-\lambda\tau s\right)/2$. In agreement with earlier 
notation \cite{muniz14}, the eigenenergies
are $E_{\boldsymbol{k}\pm}=\hbar\left(\varpi_{\boldsymbol{k}}\pm d_{\boldsymbol{k}}\right)$
where $d_{\boldsymbol{k}}=\left|\boldsymbol{d}_{\boldsymbol{k}}\right|$,
with $\left(+\right)=c$ and $\left(-\right)=v$ representing the
conduction and valence bands respectively; also, for convenience, we
denote $\omega_{cv,\boldsymbol{k}} \equiv  \hbar^{-1}\left(E_{\boldsymbol{k},c}-E_{\boldsymbol{k},v}\right)=2d_{\boldsymbol{k}}$.
The Hamiltonian is diagonalized by the unitary matrix $U_{\boldsymbol{k}}=\exp\left(-i\frac{\phi_{\boldsymbol{k}}}{2}\hat{\boldsymbol{n}}_{\boldsymbol{k}}\cdot\boldsymbol{\sigma}\right)$,
with $\hat{\boldsymbol{n}}_{\boldsymbol{k}}=\hat{\boldsymbol{z}}\times\hat{\boldsymbol{d}}_{\boldsymbol{k}}/\left|\hat{\boldsymbol{z}}\times\hat{\boldsymbol{d}}_{\boldsymbol{k}}\right|$
and $\cos\phi_{\boldsymbol{k}}=\hat{z}\cdot\hat{\boldsymbol{d}}_{\boldsymbol{k}}$.
The triad $\Xi=\left\lbrace \hat{\boldsymbol{n}}_{\boldsymbol{k}},\hat{\boldsymbol{d}}_{\boldsymbol{k}},\hat{\boldsymbol{n}}_{\boldsymbol{k}}\times\hat{\boldsymbol{d}}_{\boldsymbol{k}}\right\rbrace $
forms an orthonormal basis, so an arbitrary operator $\hat{\boldsymbol{w}}\cdot\boldsymbol{\sigma}$
can be easily written in the basis of eigenvectors $U_{\boldsymbol{k}}^{\dagger}\left(\hat{\boldsymbol{w}}\cdot\boldsymbol{\sigma}\right)U_{\boldsymbol{k}}$
by decomposing $\hat{\boldsymbol{w}}$ in the triad $\Xi$. For the
system under consideration, $\hat{\boldsymbol{n}}_{\boldsymbol{k}}=\frac{1}{k}\left(-k_{y}\hat{\boldsymbol{x}}+\tau k_{x}\hat{\boldsymbol{y}}\right)$
and $d_{\boldsymbol{k}}=\sqrt{t^{2}k^{2}+\Delta_{\tau s}^{2}}$ so
\begin{equation}
\begin{array}{rl}
\partial_{k^{b}}d_{\boldsymbol{k}}= & \frac{t^{2}k^{b}}{d_{\boldsymbol{k}}},\\
\partial_{k^{b}}\hat{\boldsymbol{d}}_{\boldsymbol{k}}= & \frac{t\left(\tau b^{x}\hat{\boldsymbol{x}}+b^{y}\hat{\boldsymbol{y}}\right)}{d_{\boldsymbol{k}}}-\frac{t^{2}k^{b}\boldsymbol{d}_{\boldsymbol{k}}}{d_{\boldsymbol{k}}^{3}}.
\end{array}
\end{equation}
The velocity operator $v_{\boldsymbol{k}}^{a}=\frac{1}{\hbar}\partial_{k^{a}}H_{\boldsymbol{k}}$
plays a fundamental role in the determination of optical properties;
written in the basis of eigenstates, it is given by 
\begin{equation}
\begin{array}{rl}
v_{\boldsymbol{k}}^{a}= & \partial_{k^{a}}\varpi_{\boldsymbol{k}}\sigma_{0}+\partial_{k^{a}}d_{\boldsymbol{k}}\sigma_{z}+d_{\boldsymbol{k}}\left(\hat{\boldsymbol{n}}_{\boldsymbol{k}}\cdot\partial_{k^{a}}\hat{\boldsymbol{d}}_{\boldsymbol{k}}\right)\hat{\boldsymbol{n}}_{\boldsymbol{k}}\cdot\sigma\\
 & +d_{\boldsymbol{k}}\left[\left(\hat{\boldsymbol{n}}_{\boldsymbol{k}}\times\hat{\boldsymbol{d}}_{\boldsymbol{k}}\right)\cdot\partial_{k^{a}}\hat{\boldsymbol{d}}_{\boldsymbol{k}}\right]\left(\hat{\boldsymbol{n}}_{\boldsymbol{k}}\times\hat{\boldsymbol{z}}\right)\cdot\sigma,
\end{array}
\end{equation}
so 
\begin{equation}
v_{cc}^{a}-v_{vv}^{a}=2\partial_{k^{a}}d_{\boldsymbol{k}}=\frac{2t^{2}k^{a}}{d_{\boldsymbol{k}}},
\end{equation}
and since 
\begin{equation}
\begin{array}{rl}
\partial_{k^{a}}\hat{\boldsymbol{d}}_{\boldsymbol{k}}\cdot\partial_{k^{b}}\hat{\boldsymbol{d}}_{\boldsymbol{k}}= & \frac{t^{2}\hat{\boldsymbol{a}}\cdot\hat{\boldsymbol{b}}}{d_{\boldsymbol{k}}^{2}}-\frac{2t^{4}k^{a}k^{b}}{d_{\boldsymbol{k}}^{4}}+\frac{t^{4}k^{a}k^{b}}{d_{\boldsymbol{k}}^{4}} \\  
= &  \frac{t^{2}\hat{\boldsymbol{a}} \cdot \hat{\boldsymbol{b}}}{d_{\boldsymbol{k}}^{2}}-\frac{t^{4}k^{a}k^{b}}{d_{\boldsymbol{k}}^{4}} , \\
\hat{\boldsymbol{d}}_{\boldsymbol{k}}\cdot\left(\partial_{k^{a}}\hat{\boldsymbol{d}}_{\boldsymbol{k}} \times \partial_{k^{b}} \hat{\boldsymbol{d}}_{\boldsymbol{k}} \right) = & \frac{t^{2}\tau\boldsymbol{d}_{\boldsymbol{k}}\cdot\left(\hat{\boldsymbol{a}}\times\hat{\boldsymbol{b}}\right)}{d_{\boldsymbol{k}}^{3}}=\frac{t^{2}\tau\Delta_{\tau s}\hat{\boldsymbol{z}}\cdot\left(\hat{\boldsymbol{a}}\times\hat{\boldsymbol{b}}\right)}{d_{\boldsymbol{k}}^{3}} , 
\end{array}
\end{equation}
then 
\begin{equation}
\begin{array}{rl}
v_{cv}^{a}v_{vc}^{b}= & d_{\boldsymbol{k}}^{2}\left[\partial_{k^{a}}\hat{\boldsymbol{d}}_{\boldsymbol{k}}\cdot\partial_{k^{b}}\hat{\boldsymbol{d}}_{\boldsymbol{k}}+i\hat{\boldsymbol{d}}_{\boldsymbol{k}}\cdot\left(\partial_{k^{a}}\hat{\boldsymbol{d}}_{\boldsymbol{k}}\times\partial_{k^{b}}\hat{\boldsymbol{d}}_{\boldsymbol{k}}\right)\right]\\
= & t^{2}\left[\hat{\boldsymbol{a}}\cdot\hat{\boldsymbol{b}}-\frac{t^{2}k^{a}k^{b}}{d_{\boldsymbol{k}}^{2}}+i\frac{\tau\Delta_{\tau s}\hat{\boldsymbol{z}}\cdot\left(\hat{\boldsymbol{a}}\times\hat{\boldsymbol{b}}\right)}{d_{\boldsymbol{k}}}\right] ,
\end{array}
\label{eq:vv}
\end{equation}
the last term in the above equation is due to the Berry curvature.

\begin{figure}[htbp!]
\begin{tabular}{lr}
\includegraphics[width=0.49\columnwidth]{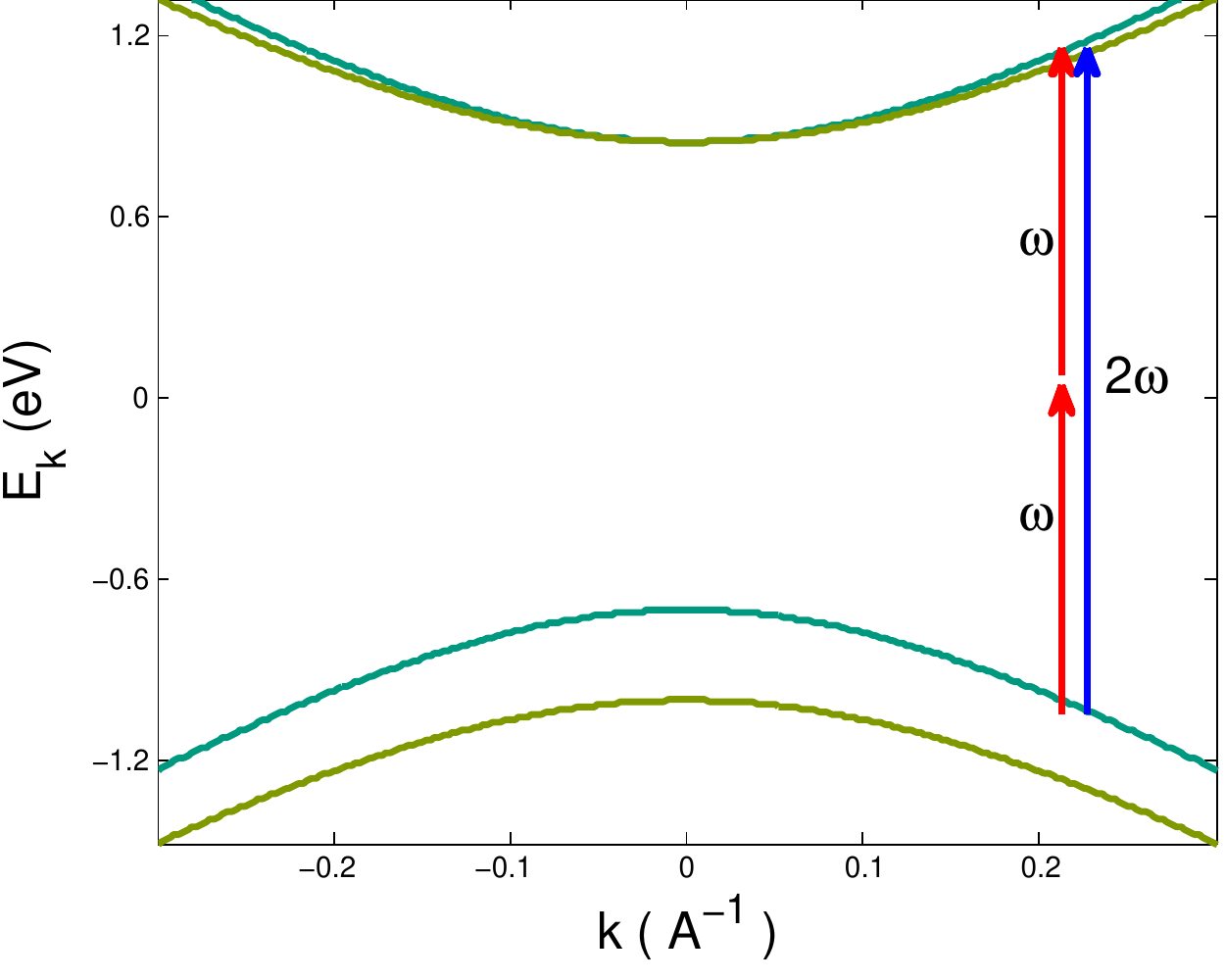}  & \includegraphics[width=0.49\columnwidth]{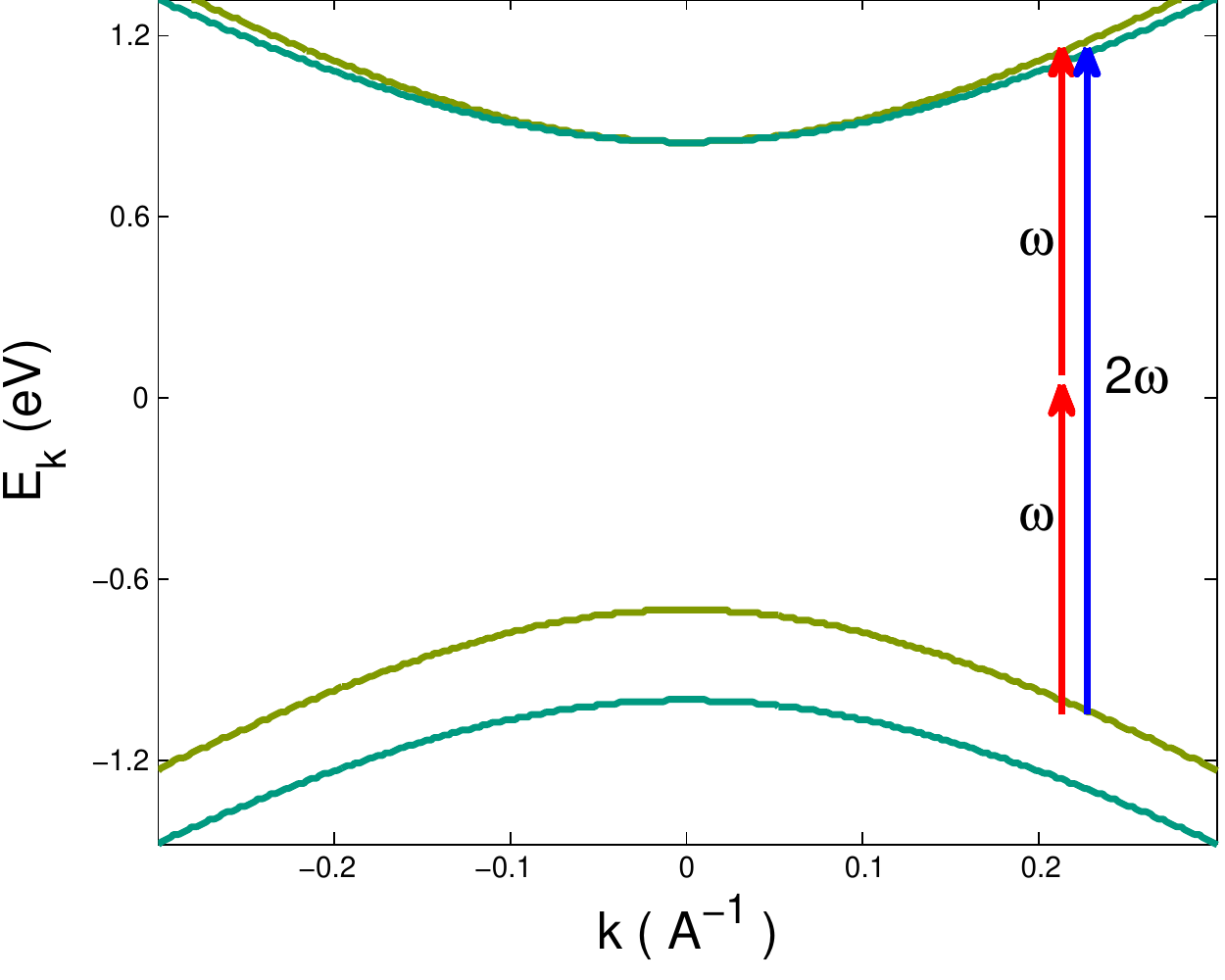} 
\end{tabular} 
\caption[]{(Color online) Bands at the two valleys with the absorption processes indicated. Different colors of the bands distinguish between spin ($\uparrow$) and ($\downarrow$). Note that bands with the same energy have opposite spins on the two different valleys.}
\label{fig:carrier} 
\end{figure}

\section{Optical injection rates }

\label{sec:inject}

The calculation for the injection rates is carried out using Fermi's Golden Rule, since it makes evident all the contributions stemming
from one- and two-photon processes and their interference; this is a feature
not shared by the Kubo formalism, for instance. The general formulation
has been already well explained in previous studies \cite{rioux12,muniz14},
so we only show the information that is specific for monolayer TMDs.

The incident light fields correspond to the vector potential $\boldsymbol{A}\left(t\right)=\sum_{n}\boldsymbol{A}\left(\omega_{n}\right)e^{-i\omega_{n}t}$,
with $\omega_{n}=\pm\omega,\pm2\omega$; the associated electric field
is given by $\boldsymbol{E}\left(t\right)=-c^{-1}\partial_{t}\boldsymbol{A}\left(t\right)$.
The injection rate for the density $\left\langle M\right\rangle $
of a quantity associated with a single-particle operator ${\cal M}=\sum_{\boldsymbol{k}}a_{\alpha,\boldsymbol{k}}^{\dagger}M_{\alpha\beta,\boldsymbol{k}}a_{\beta,\boldsymbol{k}}$,
here $\alpha$ and $\beta$ are band indices, can be decomposed into
contributions from one and two photons absorption processes with an
additional interference term $\left\langle \dot{M}\right\rangle =\left\langle \dot{M}_{1}\right\rangle +\left\langle \dot{M}_{2}\right\rangle +\left\langle \dot{M}_{i}\right\rangle $
with  
\begin{equation}
\begin{array}{rl}
\left\langle \dot{M}_{1}\right\rangle = & \underset{n=1,2}{\sum}\Lambda_{1}^{bc}\left(n\omega\right)E^{b}\left(-n\omega\right)E^{c}\left(n\omega\right),\\
\left\langle \dot{M}_{2}\right\rangle = & \Lambda_{2}^{bcde}\left(\omega\right)E^{b}\left(-\omega\right)E^{c}\left(-\omega\right)E^{d}\left(\omega\right)E^{e}\left(\omega\right),\\
\left\langle \dot{M}_{i}\right\rangle = & \Lambda_{i}^{bcd}\left(\omega\right)E^{b}\left(-\omega\right)E^{c}\left(-\omega\right)E^{d}\left(2\omega\right)+cc.
\end{array}
\end{equation}
 The optical injection coefficients $\Lambda$ associated with ${\cal M}$ are obtained through integrals over the Brillouin zone; if $\boldsymbol{k}$ is represented in polar coordinates, the integral over the radial component enforces the energy matching condition ($\omega_{cv}=\omega$ or $\omega_{cv}=2\omega$), and only the angular integral remains \cite{muniz14} 
\begin{equation}
\begin{array}{rl}
\Lambda_{1}^{bc}\left(\omega\right)= & \left.\int\frac{d\theta}{2\pi}\frac{d_{\boldsymbol{k}}\left(M_{cc,\boldsymbol{k}}-M_{vv,\boldsymbol{k}}\right)\Gamma_{1,cv}^{bc}\left(\boldsymbol{k},\omega\right)}{2t^{2}}\right|_{d_{\boldsymbol{k}}=\frac{\omega}{2}},\\
\Lambda_{2}^{bcde}\left(\omega\right)= & \left.\int\frac{d\theta}{2\pi}\frac{d_{\boldsymbol{k}}\left(M_{cc,\boldsymbol{k}}-M_{vv,\boldsymbol{k}}\right)\Gamma_{2,cv}^{bcde}\left(\boldsymbol{k},\omega\right)}{2t^{2}}\right|_{d_{\boldsymbol{k}}=\omega},\\
\Lambda_{i}^{bcd}\left(\omega\right)= & \left.\int\frac{d\theta}{2\pi}\frac{d_{\boldsymbol{k}}\left(M_{cc,\boldsymbol{k}}-M_{vv,\boldsymbol{k}}\right)\Gamma_{i,cv}^{bcd}\left(\boldsymbol{k},\omega\right)}{2t^{2}}\right|_{d_{\boldsymbol{k}}=\omega},
\end{array}\label{eq:coeff2band}
\end{equation}
where 
\begin{equation}
\begin{array}{rl}
\Gamma_{1,cv}^{bc}\left(\boldsymbol{k},\omega\right)= & \frac{e^{2}v_{cv}^{c}v_{vc}^{b}}{\hbar^{2}\omega^{2}},\\
\Gamma_{2,cv}^{bcde}\left(\boldsymbol{k},\omega\right)= & \frac{e^{4}t^{4}}{\hbar^{4}\omega^{6}}\left[\frac{k^{c}k^{e}v_{cv}^{d}v_{vc}^{b}+k^{c}k^{d}v_{cv}^{e}v_{vc}^{b}+k^{b}k^{e}v_{cv}^{d}v_{vc}^{c}+k^{b}k^{d}v_{cv}^{e}v_{vc}^{c}}{d_{\boldsymbol{k}}^{2}}\right],\\
\Gamma_{i,cv}^{bcd}\left(\boldsymbol{k},\omega\right)= & \frac{ie^{3}t^{2}}{2\hbar^{3}\omega^{4}}\left[\frac{k^{c}v_{cv}^{d}v_{vc}^{b}+k^{b}v_{cv}^{d}v_{vc}^{c}}{d_{\boldsymbol{k}}}\right],
\end{array}\label{eq:gammas2dA}
\end{equation}
 and the integrals are over the circle in the Brillouin zone set by the energy matching condition $\omega_{cv}=\omega$ or $\omega_{cv}=2\omega$.

\subsubsection*{Operators}

The quantities of interest are the densities of injected carriers
$\left\langle n\right\rangle $, spin $\left\langle S_{z}\right\rangle $
and valley $\left\langle \tau\right\rangle $ polarizations, as well
as currents of charge $\left\langle \boldsymbol{J}_{c}\right\rangle $,
spin $\left\langle \boldsymbol{J}_{S}\right\rangle $, and valley
$\left\langle \boldsymbol{J}_{\tau}\right\rangle $, which are all
computed below. The corresponding optical injection coefficients are
respectively $\xi_N$, $\zeta_N$ and $\vartheta_N$ for the densities,
and $\eta_N$, $\mu_N$ and $\nu_N$ for the currents; the subindex $N = 1,2,i$ indicates which absorption process is involved.

We keep track of the injected carriers by computing the density of
electrons injected into the conduction band. For each Dirac cone (labeled
by the indices $s$ and $\tau$), the corresponding number operator
has matrix elements $n_{cc}=1$ and $n_{vv}=0$, for all the 4 Dirac
cones. The operators corresponding to polarizations of spin $S^{z}=\frac{\hbar}{2}s\sigma_{z}$
and valley ${\cal T}=\tau\sigma_{z}$ have matrix elements $S_{cc}^{z}=\frac{\hbar}{2}s$
and $S_{vv}^{z}=-\frac{\hbar}{2}s$, and $\tau_{cc}=\tau$ and $\tau_{vv}=-\tau$,
for each Dirac cone.

The operators corresponding to currents of charge $\boldsymbol{J}_{c}=e\boldsymbol{v}$,
spin $\boldsymbol{J}_{S}=\frac{\hbar}{2}s\boldsymbol{v}$, and valley
$\boldsymbol{J}_{\tau}=\tau\boldsymbol{v}$, are expressed in terms
of the velocity operator $\boldsymbol{v}$.

Summarizing, we have 
\begin{equation}
\begin{array}{rl}
n_{cc}-n_{vv}= & 1,\\
S_{cc}^{z}-S_{vv}^{z}= & \hbar s,\\
\tau_{cc}-\tau_{vv}= & 2\tau,
\end{array}
\end{equation}
for densities, and 
\begin{equation}
\begin{array}{rl}
\boldsymbol{J}_{cc}^{c}-\boldsymbol{J}_{vv}^{c}= & e\left(\boldsymbol{v}_{cc}-\boldsymbol{v}_{vv}\right)\\
\boldsymbol{J}_{cc}^{s}-\boldsymbol{J}_{vv}^{s}= & \frac{\hbar}{2}s\left(\boldsymbol{v}_{cc}-\boldsymbol{v}_{vv}\right)\\
\boldsymbol{J}_{cc}^{\tau}-\boldsymbol{J}_{vv}^{\tau}= & \tau\left(\boldsymbol{v}_{cc}-\boldsymbol{v}_{vv}\right)
\end{array}
\end{equation}
for currents.

\subsubsection*{Optical injection coefficients}

The expressions for the various optical injection coefficients follow
from the following 
\begin{widetext}
\begin{equation}
\begin{array}{rl}
\bar{\xi}_{1,\tau s}^{bc}\left(\omega\right)= & \frac{\Theta\left(\omega-2\Delta_{\tau s}\right)e^{2}}{2\hbar^{2}\omega}\left(1+\frac{4\Delta_{\tau s}^{2}}{\omega^{2}}\right)\frac{\hat{\boldsymbol{b}}\cdot\hat{\boldsymbol{c}}}{4},\\
\bar{\xi}_{2,\tau s}^{bcde}\left(\omega\right)= & \frac{\Theta\left(\omega-\Delta_{\tau s}\right)e^{4}t^{2}}{\hbar^{4}\omega^{5}}\left(1-\frac{\Delta_{\tau s}^{2}}{\omega^{2}}\right)\left[\frac{\left(\hat{\boldsymbol{b}}\cdot\hat{\boldsymbol{d}}\right)\hat{\boldsymbol{c}}\cdot\hat{\boldsymbol{e}}+\left(\hat{\boldsymbol{b}}\cdot\hat{\boldsymbol{e}}\right)\hat{\boldsymbol{c}}\cdot\hat{\boldsymbol{d}}}{2}-2\left(1-\frac{\Delta_{\tau s}^{2}}{\omega^{2}}\right)\varphi^{bcde}\right],\\
\bar{\eta}_{i,\tau s}^{abcd}\left(\omega\right)= & \frac{i\Theta\left(\omega-\Delta_{\tau s}\right)e^{4}t^{2}}{2\hbar^{3}\omega^{3}}\left(1-\frac{\Delta_{\tau s}^{2}}{\omega^{2}}\right)\left[\frac{\left(\hat{\boldsymbol{a}}\cdot\hat{\boldsymbol{c}}\right)\hat{\boldsymbol{b}}\cdot\hat{\boldsymbol{d}}+\left(\hat{\boldsymbol{a}}\cdot\hat{\boldsymbol{b}}\right)\hat{\boldsymbol{c}}\cdot\hat{\boldsymbol{d}}}{2}-2\left(1-\frac{\Delta_{\tau s}^{2}}{\omega^{2}}\right)\varphi^{abcd}\right],
\end{array}
\end{equation}
and 
\begin{equation}
\begin{array}{rl}
\tilde{\xi}_{1,\tau s}^{bc}\left(\omega\right)= & \frac{\Theta\left(\omega-2\Delta_{\tau s}\right)e^{2}}{2\hbar^{2}\omega}\left[\frac{-\Delta_{\tau s}\hat{\boldsymbol{z}}\cdot\left(\hat{\boldsymbol{b}}\times\hat{\boldsymbol{c}}\right)}{\omega}\right],\\
\tilde{\xi}_{2,\tau s}^{bcde}\left(\omega\right)= & \frac{\Theta\left(\omega-\Delta_{\tau s}\right)e^{4}t^{2}}{\hbar^{4}\omega^{5}}\left(1-\frac{\Delta_{\tau s}^{2}}{\omega^{2}}\right)\frac{\Delta_{\tau s}}{\omega}\left[\frac{\hat{\boldsymbol{c}}\cdot\hat{\boldsymbol{e}}\left(\hat{\boldsymbol{d}}\times\hat{\boldsymbol{b}}\right)\cdot\hat{\boldsymbol{z}}+\hat{\boldsymbol{b}}\cdot\hat{\boldsymbol{e}}\left(\hat{\boldsymbol{d}}\times\hat{\boldsymbol{c}}\right)\cdot\hat{\boldsymbol{z}}+\hat{\boldsymbol{c}}\cdot\hat{\boldsymbol{d}}\left(\hat{\boldsymbol{e}}\times\hat{\boldsymbol{b}}\right)\cdot\hat{\boldsymbol{z}}+\hat{\boldsymbol{d}}\cdot\hat{\boldsymbol{b}}\left(\hat{\boldsymbol{e}}\times\hat{\boldsymbol{c}}\right)\cdot\hat{\boldsymbol{z}}}{4}\right],\\
\tilde{\eta}_{i,\tau s}^{abcd}\left(\omega\right)= & \frac{i\Theta\left(\omega-\Delta_{\tau s}\right)e^{4}t^{2}}{2\hbar^{3}\omega^{3}}\left(1-\frac{\Delta_{\tau s}^{2}}{\omega^{2}}\right)\frac{\Delta_{\tau s}}{\omega}\left[\frac{\hat{\boldsymbol{a}}\cdot\hat{\boldsymbol{c}}\left(\hat{\boldsymbol{d}}\times\hat{\boldsymbol{b}}\right)\cdot\hat{\boldsymbol{z}}+\hat{\boldsymbol{a}}\cdot\hat{\boldsymbol{b}}\left(\hat{\boldsymbol{d}}\times\hat{\boldsymbol{c}}\right)\cdot\hat{\boldsymbol{z}}}{2}\right],
\end{array}
\end{equation}
\end{widetext}
 such that  
\begin{equation}
\begin{array}{rl}
\xi_{1,\tau s}^{bc}\left(\omega\right)= & \bar{\xi}_{1,\tau s}^{bc}\left(\omega\right)+i\tau\tilde{\xi}_{1,\tau s}^{bc}\left(\omega\right),\\
\xi_{2,\tau s}^{bcde}\left(\omega\right)= & \bar{\xi}_{2,\tau s}^{bcde}\left(\omega\right)+i\tau\tilde{\xi}_{2,\tau s}^{bcde}\left(\omega\right),\\
\eta_{i,\tau s}^{abcd}\left(\omega\right)= & \bar{\eta}_{i,\tau s}^{abcd}\left(\omega\right)+i\tau\tilde{\eta}_{i,\tau s}^{abcd}\left(\omega\right),
\end{array}
\end{equation}
which are associated with carrier density and the charge current.
The others coefficients can be obtained from these by 
\begin{equation}
\begin{array}{rl}
\zeta_{\tau s}\left(\omega\right)= & \hbar s\xi_{\tau s}\left(\omega\right),\\
\vartheta_{\tau s}\left(\omega\right)= & 2\tau\xi_{\tau s}\left(\omega\right),\\
\mu_{\tau s}\left(\omega\right)= & \frac{\hbar s}{2e}\eta_{\tau s}\left(\omega\right),\\
\nu_{\tau s}\left(\omega\right)= & \frac{\tau}{e}\eta_{\tau s}\left(\omega\right).
\end{array}
\end{equation}
The results above are for one valley only, in order to find total
injected quantities it is necessary to sum the contributions from
the four valleys, with the following results for the optical injection
tensors 
\begin{equation}
\begin{array}{rl}
\xi\left(\omega\right)= & 2\left[\bar{\xi}_{++}\left(\omega\right)+\bar{\xi}_{+-}\left(\omega\right)\right],\\
\zeta\left(\omega\right)= & 2\hbar i\left[\tilde{\xi}_{++}\left(\omega\right)-\tilde{\xi}_{+-}\left(\omega\right)\right],\\
\vartheta\left(\omega\right)= & 4i\left[\tilde{\xi}_{++}\left(\omega\right)+\tilde{\xi}_{+-}\left(\omega\right)\right],
\end{array}
\end{equation}
and 
\begin{equation}
\begin{array}{rl}
\eta\left(\omega\right)= & 2\left[\bar{\eta}_{++}\left(\omega\right)+\bar{\eta}_{+-}\left(\omega\right)\right],\\
\mu\left(\omega\right)= & \frac{\hbar i}{e}\left[\tilde{\eta}_{++}\left(\omega\right)-\tilde{\eta}_{+-}\left(\omega\right)\right],\\
\nu\left(\omega\right)= & \frac{2i}{e}\left[\tilde{\eta}_{++}\left(\omega\right)+\tilde{\eta}_{+-}\left(\omega\right)\right].
\end{array}
\end{equation}
We next analyze these results for different polarizations.

\section{Results }

\label{sec:results}

For the system we are considering, one- and two-photon absorption
processes inject scalar quantities while interference processes inject
vectorial ones, so carriers, spin and valley densities are injected
by one- and two-photon absorption processes, but not from the interference
between them. Conversely, charge, spin and valley currents are injected
solely from the interference processes, not from the one- and two-photon
absorption processes.

The values of the parameters $t$, $\Delta$ and $\lambda$ used for
the plots or specific estimates are given in Table \ref{tab:param};
they correspond to the parameters of $\text{Mo}\text{S}_{2}$ \cite{xiao12}.

\begin{table}[htb]
\begin{tabular}{|c|c|c|c|c|}
\hline 
$\hbar t$  & $\hbar\lambda$  & $\hbar\Delta$  & $E_{\omega}$  & $E_{2\omega}$ \tabularnewline
\hline 
$3.5\AA\cdot eV$  & $0.15eV$  & $1.7eV$  & $4.1\cdot10^{5}\frac{V}{m}$  & $100\frac{V}{m}$ \tabularnewline
\hline 
\end{tabular}\protect\protect\caption{Values of the parameters used for the plots. }

\label{tab:param} 
\end{table}

We consider field amplitudes of $E_{\omega}=4.1\cdot10^{5}\frac{V}{m}$
for the fundamental and $E_{2\omega}=100\frac{V}{m}$ for the second
harmonic, which are indicative of the largest field intensities allowed
within the perturbative regime. These values depend on the expressions
for the injected carrier density, so we explain how they are obtained
in Sec. \ref{sec:discussion}.

\subsection{Linear polarizations }

The one- and two-photon processes do not depend on the relative orientation
of the fundamental $\boldsymbol{E}\left(\omega\right)=E_{\omega}e^{i\theta_{1}}\hat{\boldsymbol{e}}_{\omega}$
and second harmonic $\boldsymbol{E}\left(2\omega\right)=E_{2\omega}e^{i\theta_{2}}\hat{\boldsymbol{e}}_{2\omega}$
fields, where $E_{\omega}$ and $E_{2\omega}$ are real. Therefore
we show here the results for the injection coefficients $\Lambda_{1}$
and $\Lambda_{2}$, while the results for $\Lambda_{i}$ are displayed
for the special cases of parallel and perpendicular polarizations.

The carrier density injection rate is given by 
\begin{equation}
\left\langle \dot{n}\right\rangle =\bar{\xi}_{1}^{xx}\left(2\omega\right)E_{2\omega}^{2}+\bar{\xi}_{2}^{xxxx}\left(\omega\right)E_{\omega}^{4}.
\end{equation}
The injection rates of of spin and valley current vanish for linear
polarizations. The one- and two-photon injection rates of the currents
vanish as well $\left\langle \dot{\boldsymbol{J}}_{1}\right\rangle =\left\langle \dot{\boldsymbol{J}}_{2}\right\rangle =0$,
which are injected only through the interference process.

\subsubsection*{Parallel orientations }

Only the interference processes depend on the relative orientation
of $\boldsymbol{E}\left(\omega\right)=E_{\omega}e^{i\theta_{1}}\hat{\boldsymbol{e}}_{\omega}$
and $\boldsymbol{E}\left(2\omega\right)=E_{2\omega}e^{i\theta_{2}}\hat{\boldsymbol{e}}_{\omega}$.
The relative phase parameter is $\Delta\theta=\theta_{2}-2\theta_{1}$.

The charge current injection rate is given by 
\begin{equation}
\left\langle \dot{\boldsymbol{J}}_{c}\right\rangle =2\hat{\boldsymbol{e}}_{\omega}i\bar{\eta}_{i}^{xxxx}\left(\omega\right)\sin\left(\Delta\theta\right)E_{\omega}^{2}E_{2\omega}.
\end{equation}
The spin and valley currents vanish for linearly polarized light.

The direction of the polarization vector provides control of the angle
of the injected current, while the relative phase parameter of the
light beams can control only their magnitude and orientation.

\subsubsection*{Perpendicular orientations }

Here we have $\boldsymbol{E}\left(\omega\right)=E_{\omega}e^{i\theta_{1}}\hat{\boldsymbol{e}}_{\omega}$
and $\boldsymbol{E}\left(2\omega\right)=E_{2\omega}e^{i\theta_{2}}\hat{\boldsymbol{e}}_{2\omega}$
with $\hat{\boldsymbol{e}}_{2\omega}=\hat{\boldsymbol{z}}\times\hat{\boldsymbol{e}}_{\omega}$.
The relative phase parameter is again $\Delta\theta=\theta_{2}-2\theta_{1}$.

The charge current injection rate is given by 
\begin{equation}
\left\langle \dot{\boldsymbol{J}}^{c}\right\rangle =2\hat{\boldsymbol{e}}_{2\omega}i\bar{\eta}_{i}^{yxxy}\left(\omega\right)\sin\left(\Delta\theta\right)E_{\omega}^{2}E_{2\omega},
\end{equation}
and the rates for spin and valley currents are 
\begin{equation}
\begin{array}{rl}
\left\langle \dot{\boldsymbol{J}}^{s}\right\rangle = & 2\hat{\boldsymbol{e}}_{\omega}\tilde{\mu}_{i}^{xxxy}\left(\omega\right)\cos\left(\Delta\theta\right)E_{\omega}^{2}E_{2\omega},\\
\left\langle \dot{\boldsymbol{J}}^{\tau}\right\rangle = & 2\hat{\boldsymbol{e}}_{\omega}\tilde{\nu}_{i}^{xxxy}\left(\omega\right)\cos\left(\Delta\theta\right)E_{\omega}^{2}E_{2\omega}.
\end{array}
\end{equation}
The charge current is injected along the direction of the second harmonic
field while the spin and valley currents are injected along the direction
of the fundamental field. The relative phase parameter $\Delta \theta$ controls their magnitude, favoring either charge or spin and valley currents.

\begin{figure*}[bthp!]
\begin{tabular}{rrr}
\includegraphics[width=0.33\textwidth]{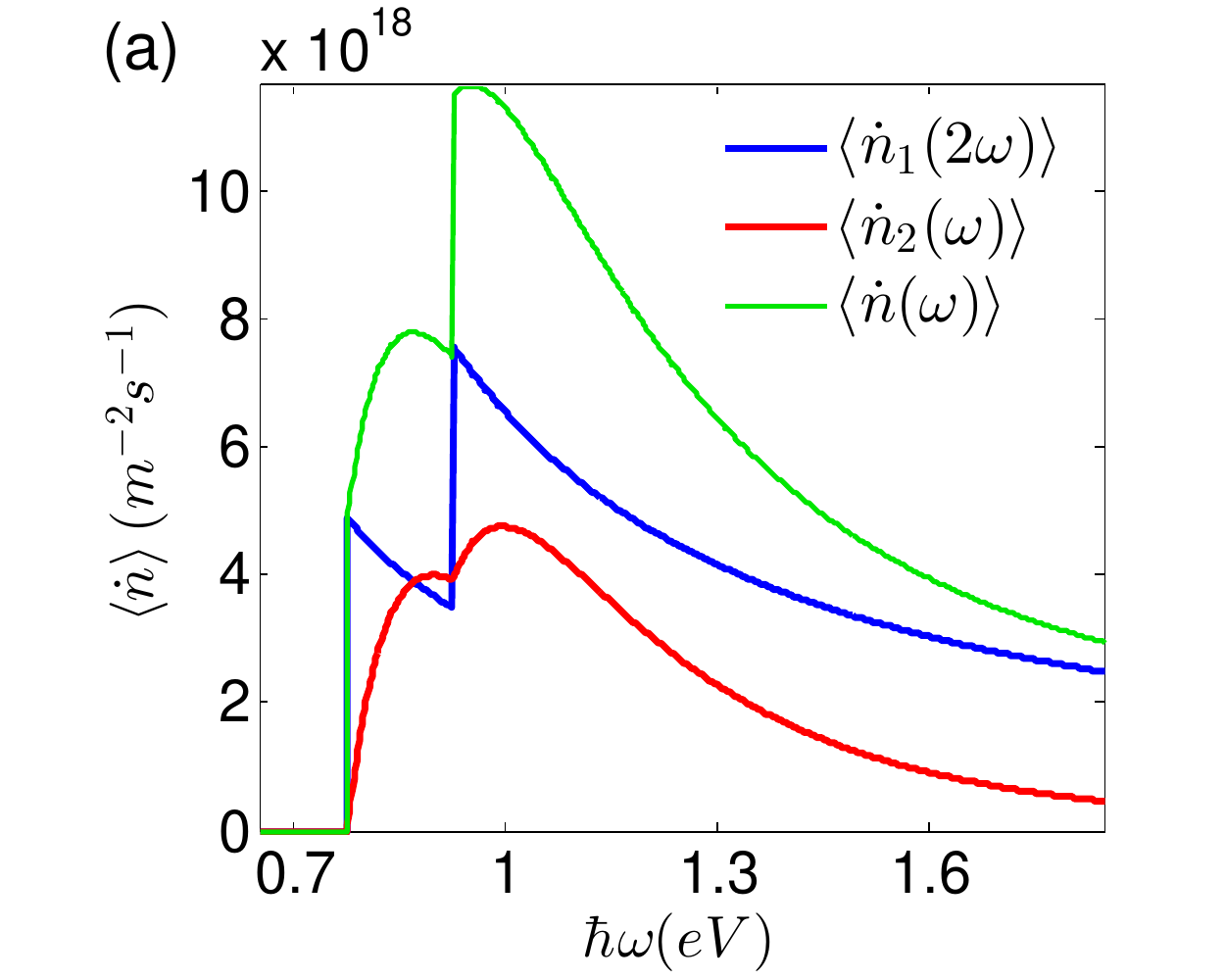}  & \includegraphics[width=0.33\textwidth]{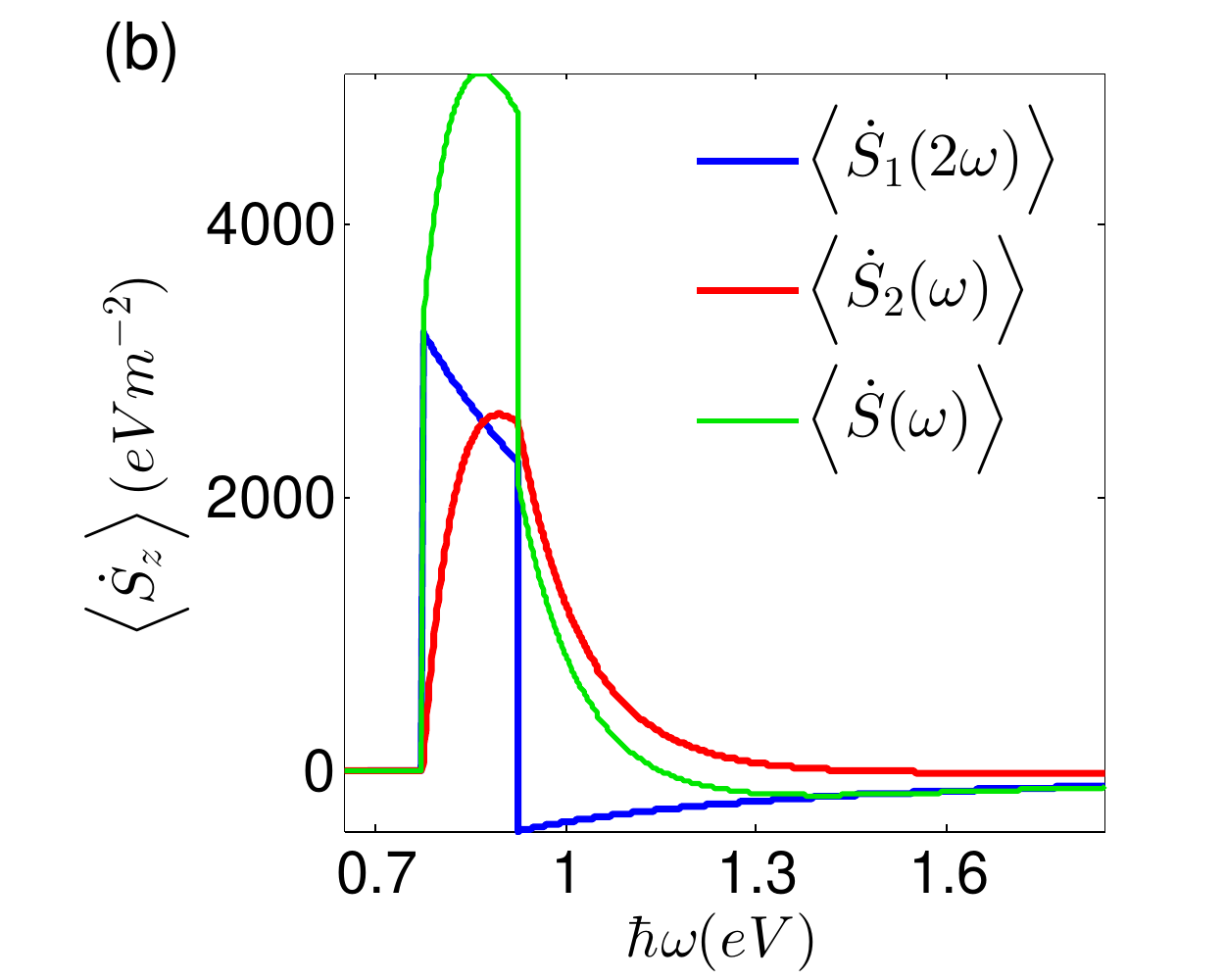}  & \includegraphics[width=0.33\textwidth]{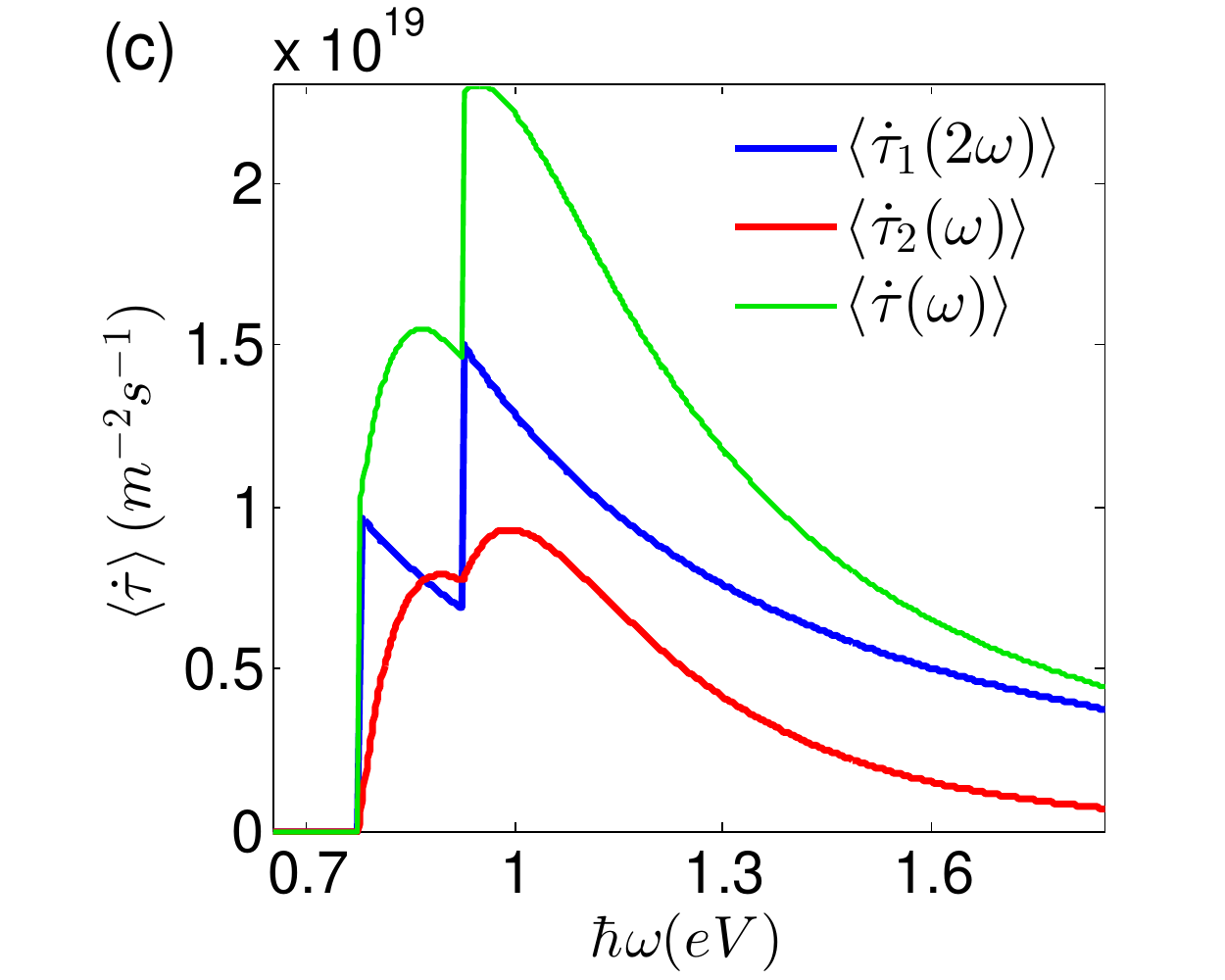} 
\end{tabular}
\caption[]{(Color online) Injection rates for (a) carrier, (b) spin, and (c) valley densities. }
\label{fig:density} 
\end{figure*}

\begin{figure*}[tbhp!]
\begin{tabular}{rrr}
\includegraphics[width=0.33\textwidth]{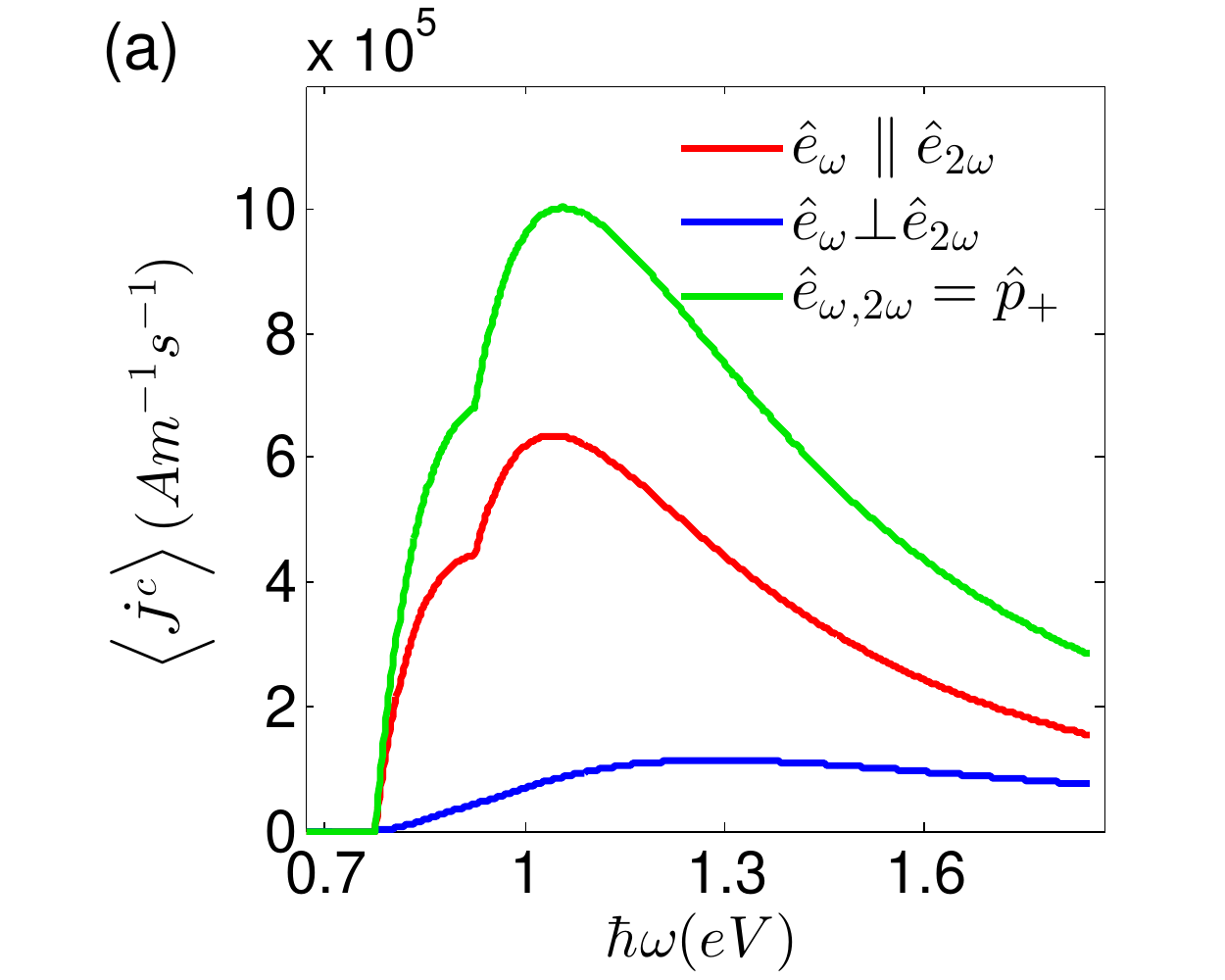}  & \includegraphics[width=0.33\textwidth]{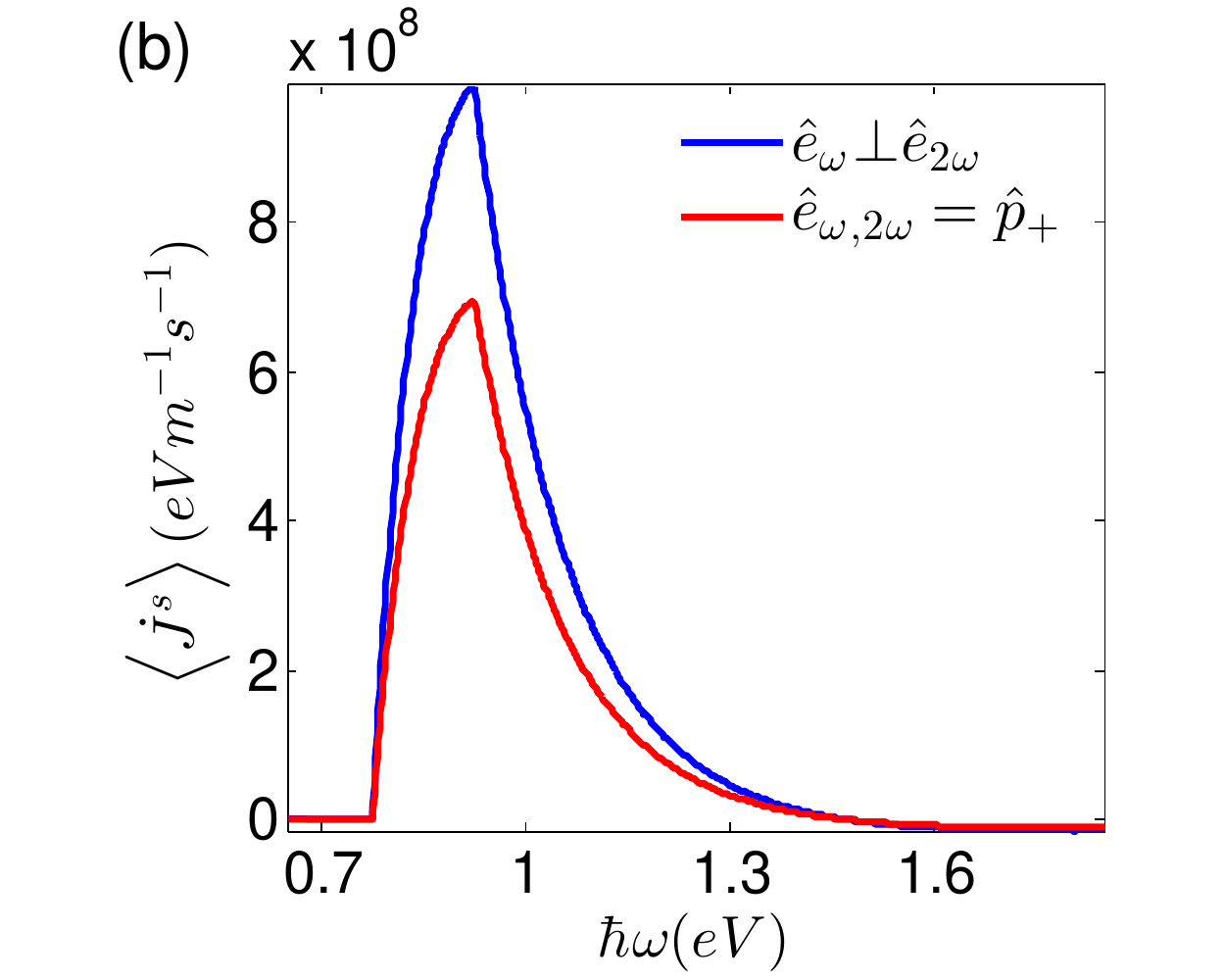}  & \includegraphics[width=0.33\textwidth]{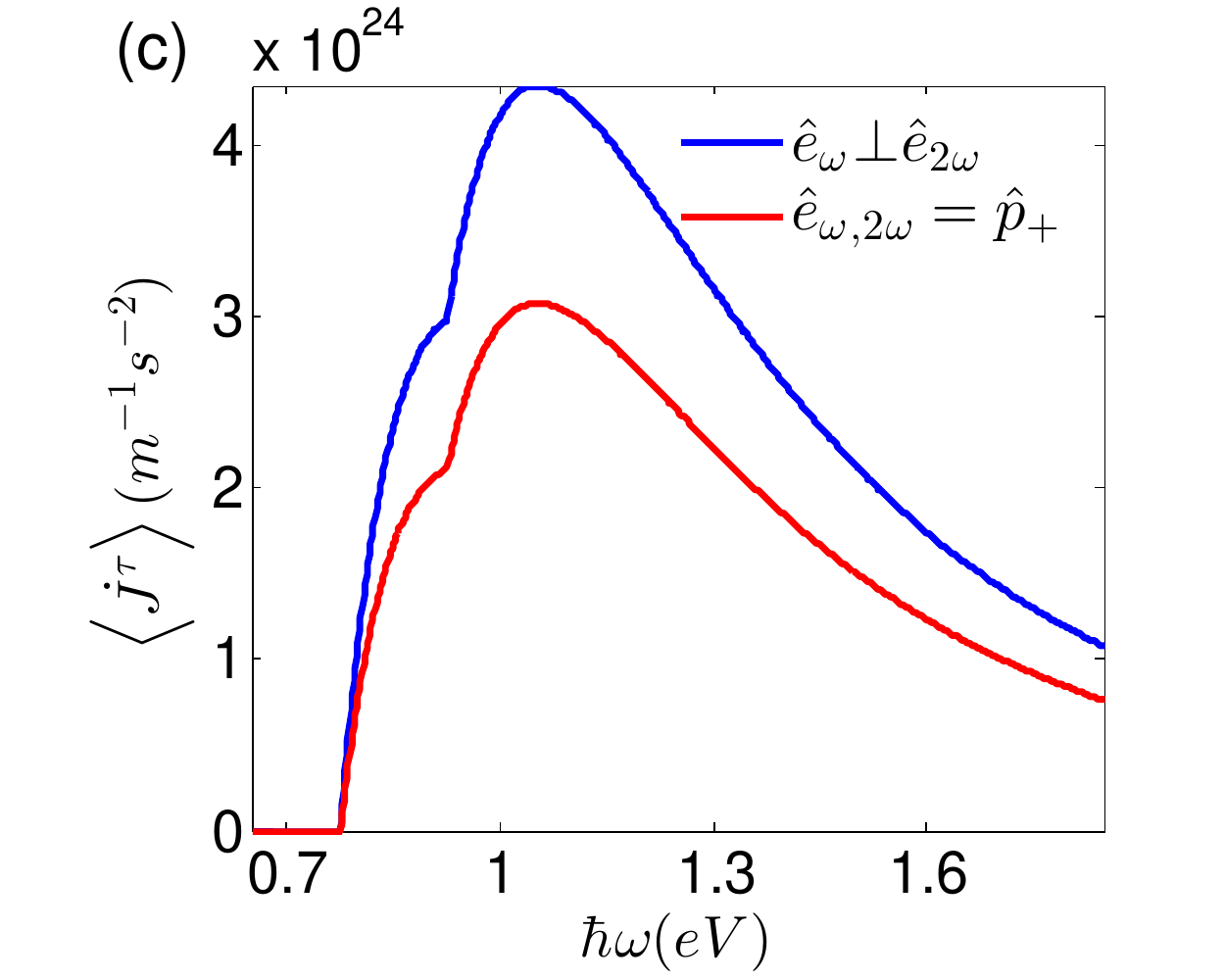} 
\end{tabular}
\caption[]{(Color online) Injection rates for current densities of (a) charge,
(b) spin, and (c) valley. }
\label{fig:current} 
\end{figure*}

\subsection{Circular polarizations }

For circular polarizations $\boldsymbol{E}\left(\omega\right)=E_{\omega}e^{i\theta_{1}}\hat{\boldsymbol{p}}_{h_{1}}$
and $\boldsymbol{E}\left(2\omega\right)=E_{2\omega}e^{i\theta_{2}}\hat{\boldsymbol{p}}_{h_{2}}$
where $h_{1},h_{2}=\pm1$ are the helicities of the light fields propagating along the $\hat{z}$ direction,
and $\hat{\boldsymbol{p}}_{\pm}=\left(\hat{\boldsymbol{x}}\pm i\hat{\boldsymbol{y}}\right)/\sqrt{2}$,
so $\hat{\boldsymbol{p}}_{h}\cdot\hat{\boldsymbol{p}}_{h}=0$ and
$\hat{\boldsymbol{p}}_{+}\cdot\hat{\boldsymbol{p}}_{-}=1$ as well
as $\hat{\boldsymbol{p}}_{-}\times\hat{\boldsymbol{p}}_{+}=i\hat{\boldsymbol{z}}$.
The relative phase parameter is still $\Delta\theta=\theta_{2}-2\theta_{1}$.

The injection rates for densities are given by 
\begin{equation}
\begin{array}{rl}
\left\langle \dot{n}\right\rangle = & \bar{\xi}_{1}^{-+}\left(2\omega\right)E_{2\omega}^{2}+\bar{\xi}_{2}^{--++}\left(\omega\right)E_{\omega}^{4},\\
\left\langle \dot{S}_{z}\right\rangle = & h_{2}\tilde{\zeta}_{1}^{-+}\left(2\omega\right)E_{2\omega}^{2}+h_{1}\tilde{\zeta}_{2}^{--++}\left(\omega\right)E_{\omega}^{4},\\
\left\langle \dot{\tau}\right\rangle = & h_{2}\tilde{\vartheta}_{1}^{-+}\left(2\omega\right)E_{2\omega}^{2}+h_{1}\tilde{\vartheta}_{2}^{--++}\left(\omega\right)E_{\omega}^{4}.
\end{array}
\end{equation}
The spin and valley density injection rates depend directly on the
Berry curvature, while the carrier density injection is independent
of it. As expected, the orientation of the spin and valley polarizations
are set by the helicities of the incident fields.

\textit{Equal helicities} - The interference process depends on the
relative helicities of the two light fields. And there are contributions
to the currents only when the helicities are equal, $\boldsymbol{E}\left(\omega\right)=E_{\omega}e^{i\theta_{1}}\hat{\boldsymbol{p}}_{h}$
and $\boldsymbol{E}\left(2\omega\right)=E_{2\omega}e^{i\theta_{2}}\hat{\boldsymbol{p}}_{h}$.

The charge current injection rate is given by 
\begin{equation}
\left\langle \dot{\boldsymbol{J}}^{c}\right\rangle =\sqrt{2}\left[\hat{\boldsymbol{x}}\sin\left(\Delta\theta\right)+\hat{\boldsymbol{y}}\cos\left(\Delta\theta\right)\right]i\bar{\eta}_{i}^{+--+}\left(\omega\right)E_{\omega}^{2}E_{2\omega},
\end{equation}
and the rates for spin and valley currents are 
\begin{equation}
\begin{array}{rl}
\left\langle \dot{\boldsymbol{J}}^{s}\right\rangle = & \sqrt{2}h\left[\hat{\boldsymbol{x}}\sin\left(\Delta\theta\right)+\hat{\boldsymbol{y}}\cos\left(\Delta\theta\right)\right]i\tilde{\mu}_{i}^{+--+}\left(\omega\right)E_{\omega}^{2}E_{2\omega},\\
\left\langle \dot{\boldsymbol{J}}^{\tau}\right\rangle = & \sqrt{2}h\left[\hat{\boldsymbol{x}}\sin\left(\Delta\theta\right)+\hat{\boldsymbol{y}}\cos\left(\Delta\theta\right)\right]i\tilde{\nu}_{i}^{+--+}\left(\omega\right)E_{\omega}^{2}E_{2\omega}.
\end{array}
\end{equation}
All the currents are now injected along the same direction, which
can be controlled by the relative phase between the light fields.
The orientation of the spin and valley currents are set by the helicity
of the incident fields. The spin and valley density injection rates
depend directly on the Berry curvature, while the carrier density
injection is independent of it.

\textit{Opposite helicities} - When the light fields have different
helicities, the injection rates from interference vanish for all the
currents of interest.

\section{Discussion }

\label{sec:discussion}

The validity of our calculations for the optical injection rates depends on the validity of the perturbative regime, which requires  
that the fraction of the injected carrier population relative to the
total number of states in the range of energies covered by the laser
pulse be small \cite{muniz14}. The duration of the pulse ${\cal T}$ sets
the frequency broadening of the laser $\Delta\omega=\frac{2\pi}{{\cal T}}$,
which in turn - via the dispersion relation- determines the area $a$ of
the Brillouin zone that can be populated by carriers, $a=2\pi k\Delta k$.
The momentum width $\Delta k$ is set by the dispersion relation and
is proportional to $\Delta\omega$. The number of states available
in this area of the Brillouin zone is $a/a_{1}$, where $a_{1}=\frac{\left(2\pi\right)^{2}}{L^{2}}$
is the area occupied by one state. The maximum amplitudes of the laser
fields are restricted by the condition that the number of injected
carriers with additional energy $2\hbar\omega$ is at most $5\%$
of the total number of carrier states in the allowed energy range
\begin{equation}
\left(\bar{\xi}_{1}^{-+}\left(2\omega\right)E_{2\omega}^{2}+\bar{\xi}_{2}^{--++}\left(\omega\right)E_{\omega}^{4}\right){\cal T}L^{2}<0.05\frac{a}{a_{1}}.\label{eq:condition}
\end{equation}
We then estimate the amplitudes by imposing the additional condition
$\bar{\xi}_{1}^{-+}\left(2\omega\right)E_{2\omega}^{2}=\bar{\xi}_{2}^{--++}\left(\omega\right)E_{\omega}^{4}$,
which gives optimal interference between the absorption processes
\cite{rioux12}. For pulses lasting $1ns$ with a frequency corresponding
to $\hbar\omega=0.9eV$, the field amplitudes found are $E_{\omega}=4.1\cdot10^{5}\frac{V}{m}$
for the fundamental and $E_{2\omega}=100\frac{V}{m}$ for the second
harmonic, which correspond to laser intensities of $22\frac{kW}{cm^{2}}$
and $1.3\frac{mW}{cm^{2}}$, respectively. We use these values for
all $\hbar\omega$ in Figs. \ref{fig:density} and \ref{fig:current}.
Although Eq. \eqref{eq:condition} is only satisfied for $\hbar\omega=0.9eV$,
it can be used to determine the appropriate field amplitudes for other frequencies. 

Although the detection of the spin or valley polarized current is difficult, it can be done by pump-probe experiments \cite{driel06prl, driel06ssc} with circularly polarized light. This allows for measuring the separation
between the two components of spin or valley after the current is injected. Experiments using an analogous technique have already been performed for monolayer TMDs \cite{sallen12, song13}.

Corrections to the injection rate coefficients due to the electron-hole interaction lead to a shift $\delta$ in the phase parameter $\Delta \theta$, which becomes $\Delta \theta = \theta_2 - 2 \theta_1 +\delta $. For semiconductors, this shift is too small \cite{bhat05}. However, since calculations for exciton binding energies \cite{louie13} indicate that monolayer TMDs have a stronger electron-hole interaction, it is reasonable to expect a considerable shift $\delta$ for them. This phase shift can be measured by simply varying the relative phases of the incident fields, and measuring the phases that lead to the maximum injection rates since they are proportional to either $\sin\left(\Delta\theta\right)$ or $\cos\left(\Delta\theta\right)$. 
Hence the experiments we are suggesting here could serve as a probe of the electron-hole interactions in these materials.  

Finally, we emphasize the advantage of the all-optical method in controlling the injected currents. The results of our calculations show that it is possible to control the direction and intensity of the injected currents by simply changing the relative phase of the fields, which can be achieved in a time scale limited only by the duration of the pulses used. 
This is perhaps more dramatic when the perpendicular linear polarizations are considered. In this case, the phase parameter $\Delta \theta$ allows to select between charge currents or perpendicular spin and valley currents. A similar effect usually occurs when DC fields are used for photocurrent injection \cite{xu14}, where a charge current is converted into perpendicular spin and valley currents, due to the opposite Berry curvature in the two valleys. But in that process it is not possible to control the currents, or fast-switch between the two cases, in contrast to the all-optical method considered in this paper. 
We therefore expect that our results will be helpful for understanding the details of these promising materials, and clarifying their potential to implement ultra-fast optical switching.

\acknowledgments

We thank Andor Kormanyos and Jin-Luo Cheng for helpful discussions.
This work was supported by the Natural Sciences and Engineering Research Council of Canada (NSERC).

\appendix

\section{Optical injection coefficients for linear and circular polarizations }

\label{app:coeffs}

The coefficients used for one- and two-photon absorption processes, obtained from Eq. \ref{eq:coeff2band}, are 
\begin{equation}
\begin{array}{rl}
\bar{\xi}_{1,\tau s}^{xx}\left(\omega\right)= & \frac{\Theta\left(\omega-2\Delta_{\tau s}\right)e^{2}}{8\hbar^{2}\omega}\left(1+\frac{4\Delta_{\tau s}^{2}}{\omega^{2}}\right),\\
\bar{\xi}_{2,\tau s}^{xxxx}\left(\omega\right)= & \frac{\Theta\left(\omega-\Delta_{\tau s}\right)e^{4}t^{2}}{4\hbar^{4}\omega^{5}}\left(1-\frac{\Delta_{\tau s}^{2}}{\omega^{2}}\right)\left(1+\frac{3\Delta_{\tau s}^{2}}{\omega^{2}}\right),
\end{array}
\end{equation}
and for linear polarization.

For circular polarization we have 
\begin{equation}
\begin{array}{rl}
\bar{\xi}_{1,\tau s}^{-+}\left(\omega\right)= & \frac{\Theta\left(\omega-2\Delta_{\tau s}\right)e^{2}}{8\hbar^{2}\omega}\left(1+\frac{4\Delta_{\tau s}^{2}}{\omega^{2}}\right),\\
\bar{\xi}_{2,\tau s}^{--++}\left(\omega\right)= & \frac{\Theta\left(\omega-\Delta_{\tau s}\right)e^{4}t^{2}}{2\hbar^{4}\omega^{5}}\left(1-\frac{\Delta_{\tau s}^{2}}{\omega^{2}}\right)\left(1+\frac{\Delta_{\tau s}^{2}}{\omega^{2}}\right),
\end{array}
\end{equation}
and 
\begin{equation}
\begin{array}{rl}
\tilde{\xi}_{1,\tau s}^{-+}\left(\omega\right)= & \frac{-i\Theta\left(\omega-2\Delta_{\tau s}\right)e^{2}}{2\hbar^{2}\omega}\left(\frac{\Delta_{\tau s}}{\omega}\right),\\
\tilde{\xi}_{2,\tau s}^{--++}\left(\omega\right)= & \frac{-i\Theta\left(\omega-\Delta_{\tau s}\right)e^{4}t^{2}}{\hbar^{4}\omega^{5}}\left(1-\frac{\Delta_{\tau s}^{2}}{\omega^{2}}\right)\frac{\Delta_{\tau s}}{\omega}.
\end{array}
\end{equation}
The interference coefficients for linear polarizations are 
\begin{equation}
\begin{array}{rl}
\bar{\eta}_{i,\tau s}^{xxxx}\left(\omega\right)= & \frac{i\Theta\left(\omega-\Delta_{\tau s}\right)e^{4}t^{2}}{8\hbar^{3}\omega^{3}}\left(1-\frac{\Delta_{\tau s}^{2}}{\omega^{2}}\right)\left(1+\frac{3\Delta_{\tau s}^{2}}{\omega^{2}}\right),\\
\bar{\eta}_{i,\tau s}^{yxxy}\left(\omega\right)= & \frac{-i\Theta\left(\omega-\Delta_{\tau s}\right)e^{4}t^{2}}{8\hbar^{3}\omega^{3}}\left(1-\frac{\Delta_{\tau s}^{2}}{\omega^{2}}\right)^{2}
\end{array}
\end{equation}
and 
\begin{equation}
\begin{array}{rl}
\tilde{\eta}_{i,\tau s}^{xxxy}\left(\omega\right)= & \frac{-i\Theta\left(\omega-\Delta_{\tau s}\right)e^{4}t^{2}}{2\hbar^{3}\omega^{3}}\left(1-\frac{\Delta_{\tau s}^{2}}{\omega^{2}}\right)\frac{\Delta_{\tau s}}{\omega},\end{array}
\end{equation}
while for circular polarizations the coefficients are 
\begin{equation}
\begin{array}{rl}
\bar{\eta}_{i,\tau s}^{+--+}\left(\omega\right)= & \frac{i\Theta\left(\omega-\Delta_{\tau s}\right)e^{4}t^{2}}{4\hbar^{3}\omega^{3}}\left(1-\frac{\Delta_{\tau s}^{2}}{\omega^{2}}\right)\left(1+\frac{\Delta_{\tau s}^{2}}{\omega^{2}}\right),\\
\tilde{\eta}_{i,\tau s}^{+--+}\left(\omega\right)= & \frac{\Theta\left(\omega-\Delta_{\tau s}\right)e^{4}t^{2}}{2\hbar^{3}\omega^{3}}\left(1-\frac{\Delta_{\tau s}^{2}}{\omega^{2}}\right)\frac{\Delta_{\tau s}}{\omega}.
\end{array}
\end{equation}
The other injection rate coefficients are obtained from the ones above.

\end{document}